\documentclass[%
 reprint,
 superscriptaddress,
 amsmath,amssymb,
 aps,
 showkeys,
]{revtex4-2}

\usepackage{graphicx}%
\usepackage{dcolumn}%
\usepackage{bm}%
\usepackage{multirow}
\usepackage{booktabs}
\usepackage{float}
\usepackage{placeins}
\usepackage{xcolor} %
\usepackage[colorlinks=true,linkcolor=blue,citecolor=blue,urlcolor=blue]{hyperref}%
\usepackage{miller}

\begin{document}

\preprint{APS/123-QED}

\title{
Unveiling the mechanisms of motion of synchro-Shockley dislocations in Laves phases\\
}

\author{Zhuocheng Xie}
\affiliation{Institute of Physical Metallurgy and Materials Physics, RWTH Aachen University, 52056 Aachen, Germany}

\author{Dimitri Chauraud}
\affiliation{Max-Planck-Institut für Eisenforschung GmbH, Max-Planck-Str. 1, 40237 Düsseldorf, Germany}

\author{Achraf Atila}
\affiliation{Max-Planck-Institut für Eisenforschung GmbH, Max-Planck-Str. 1, 40237 Düsseldorf, Germany}
\affiliation{Department of Materials Science and Engineering, Institute I: General Materials Properties, Friedrich-Alexander-Universität Erlangen-Nürnberg, 91058 Erlangen, Germany}

\author{Erik Bitzek}
\affiliation{Max-Planck-Institut für Eisenforschung GmbH, Max-Planck-Str. 1, 40237 Düsseldorf, Germany}
\affiliation{Department of Materials Science and Engineering, Institute I: General Materials Properties, Friedrich-Alexander-Universität Erlangen-Nürnberg, 91058 Erlangen, Germany}

\author{Sandra Korte-Kerzel}
\affiliation{Institute of Physical Metallurgy and Materials Physics, RWTH Aachen University, 52056 Aachen, Germany}

\author{Julien Guénolé}
\email[]{julien.guenole@univ-lorraine.fr}
\affiliation{Université de Lorraine, CNRS, Arts et Métiers ParisTech, LEM3, 57070 Metz, France}
\affiliation{Labex Damas, Université de Lorraine, 57070 Metz, France}

\date{\today}

\begin{abstract}
In Laves phases, synchroshear is the dominant basal slip mechanism. It is accomplished by the glide of synchro-Shockley dislocations. However, the atomic-scale mechanisms of motion of such zonal dislocations are still not well understood. In this work, using atomistic simulations, two 30\textdegree{} synchro-Shockley dislocations with different Burgers vectors and core structures and energies are identified. We demonstrate that nucleation and propagation of kink pairs is the energetically favorable mechanism for the motion of the synchro-Shockley dislocation (partial I). Vacancy hopping and interstitial shuffling are identified as two key mechanisms related to kink propagation and we investigated how vacancies and antisite defects assist kink nucleation and propagation, which is crucial for kink mobility. Additionally, we identified a mechanism of non-sequential atomic shuffling for the motion of the synchro-Shockley dislocation (partial II). These findings provide insights into the dependency on temperature and chemical composition of plastic deformation induced by zonal dislocations in Laves phases and the many related topologically close-packed phases.
\end{abstract}


\maketitle

\section{\label{Intro}Introduction}

Laves phases are topologically close-packed structures that form in many alloys and have a large impact on their mechanical properties due to the high strength compared to the matrix phases \cite{sinha1972topologically,paufler2011early,stein2021laves}. Laves phase alloys often exhibit excellent mechanical properties at high temperatures, however, their extreme brittleness at ambient temperatures limits their applications as structural materials \cite{livingston1992laves,pollock2010weight,stein2021laves}. The understanding of the underlying deformation mechanisms of Laves phases is thus crucial for tailoring material properties of the composites.

Laves phases with the ideal chemical composition AB\textsubscript{2} have three common polytypes: cubic MgCu\textsubscript{2} (C15), hexagonal MgZn\textsubscript{2} (C14) and MgNi\textsubscript{2} (C36). Laves crystals have a layered structure along the basal or $\{ 1 1 1 \}$ planes, which consists of quadruple atomic layers. 
The quadruple atomic layers in turn consist of a single-layer of B-type atoms forming a Kagomé net and a triple-layer with an A–B–A structure. The same layers also forms part of related structures as an intergrowth with other structural elements, such as CaCu$_5$ and Zr$_4$Al$_3$, the latter forming the $\mu$-phases \cite{PartheGmelin1993Handbook,schroders2019structure}. 

Synchroshear, as a dominant mechanism for dislocation-mediated plasticity on the basal plane in Laves phases, was already predicted in the 1960s \cite{kramer1968gittergeometrische}.
It was later confirmed by experimental observations of synchro-Shockley dislocations and synchroshear-induced stacking faults in the C14 HfCr\textsubscript{2} Laves phase \cite{chisholm2005dislocations}. Recently, $ab$ $initio$ calculations \cite{vedmedenko2008first} and atomistic simulations using semi-empirical potentials \cite{guenole2019basal} confirmed synchroshear as the energetically favorable mechanism compared to other crystallographic slip mechanisms for basal slip in Laves phases.
Synchro-Shockley dislocations, a type of zonal dislocation \cite{kronberg1957plastic, anderson2017theory}, are formed by the cooperative motion of two coupled Shockley partial dislocations on the adjacent planes of a triple-layer \cite{hazzledine1992synchroshear}. After the glide of a synchro-Shockley dislocation in a C14 (or C15) Laves phase, the alternate triple-layer transforms into a slab of C15 (or C14) structure, thus forming a stacking fault. 
I.e., synchroshear in Laves phases is always associated with the creation and extension of a stacking fault. 

In Laves phases, point defects such as anti-site atoms and vacancies widely exist in off-stoichiometric compositions and at high temperatures \cite{zhu1999point,stein2021laves}. The presence of these point defects has significant effects on the deformation behavior, such as hardness \cite{voss2008composition,takata2016nanoindentation,luo2020composition} and phase transformation kinetics \cite{kumar2004polytypic}. A progressive decrease in hardness at B-type-rich off-stoichiometric compositions was reported in nanoindentation experiments on NbCo\textsubscript{2} \cite{voss2008composition,luo2020composition} and NbFe\textsubscript{2} \cite{voss2008composition,takata2016nanoindentation} Laves phases. Single-phase NbCr\textsubscript{2} exhibits more rapid synchroshear-induced phase transformation than TiCr\textsubscript{2} and TaCr\textsubscript{2} counterparts, and the transformation is rendered sluggish by the addition of substitutional defects \cite{kumar2004polytypic}. These experimental observations were attributed to the interactions between synchro-Shockley dislocations with constitutional and thermal point defects affecting the dislocation mobility \cite{kumar2004polytypic,takata2016nanoindentation}. 

Although existence and the geometry of slip by synchroshear is well established, the atomic-scale mechanisms of motion of synchro-Shockley dislocations on the basal plane in Laves phases are still not well understood. Kink propagation and short-range diffusion were proposed as possible mechanisms of dislocation motion in Laves phases in the 1990s \cite{hazzledine1992synchroshear}, however, there has been so far no evidence from experiments or modelling. In addition, the effects of  point defects on dislocation motion in Laves phases are yet to be explored.

In this study, the core structures and energies of synchro-Shockley dislocations in C14 and C15 Laves phases were investigated using atomistic simulations. The mechanisms of motion of synchro-Shockley dislocations with and without point defects and corresponding activation energies were determined by identifying transition states on the potential energy surfaces. As stacking fault has been confirmed as the dominant defect structure on the basal slip plane in Laves phases instead of perfect dislocation \cite{hazzledine1992synchroshear,chisholm2005dislocations}, the motion of synchro-Shockley dislocations was aligned to the direction of expansion of the stacking fault.

\section{\label{Methods}Simulation methods}

The atomistic simulations presented in this study were performed using the MD software package LAMMPS \cite{LAMMPS}. 
The interatomic interactions were modeled by the modified embedded atom method (MEAM) potential by Kim et al. \cite{kim2015modified} for Mg-Ca, the MEAM potential by Jang et al. \cite{jang2021modified} was used for the Al-Ca system. 
Both potentials reasonably describe the mechanical properties of C14 CaMg\textsubscript{2} and C15 CaAl\textsubscript{2} Laves phases as compared to experiments and $ab$ $initio$ calculations (see TABLE S I in \cite{SuppMat}), e.g., the Kim potential shows a close match of the basal stacking fault energy of C14 CaMg\textsubscript{2} compared to the density functional theory (DFT) value. Additionally, both potentials successfully predicted the synchroshear as the energetically favorable mechanism for propagating dislocations within the basal and $\{ 1 1 1 \}$ planes in C14 CaMg\textsubscript{2} \cite{guenole2019basal} and C15 CaAl\textsubscript{2} Laves phases, respectively.

The C14 CaMg\textsubscript{2} and C15 CaAl\textsubscript{2} Laves structures were constructed using Atomsk \cite{hirel2015atomsk} with the lattice constant $a_{0}$ of the respective Laves phase at 0~K
\cite{kim2015modified,jang2021modified} and the following crystallographic orientations: for C14 $\mathbf{x}=[ 1 1 \bar{2} 0 ]$, $\mathbf{y}=[ \bar{1} 1 0 0 ]$ and $\mathbf{z}=[ 0 0 0 1 ]$; for C15 $\mathbf{x}=[ 1 \bar{1} 0 ]$, $\mathbf{y}=[ 1 1 \bar{2} ]$ and $\mathbf{z}=[ 1 1 1 ]$. 

\begin{figure*}[htbp!]
\centering
\includegraphics[width=\textwidth]{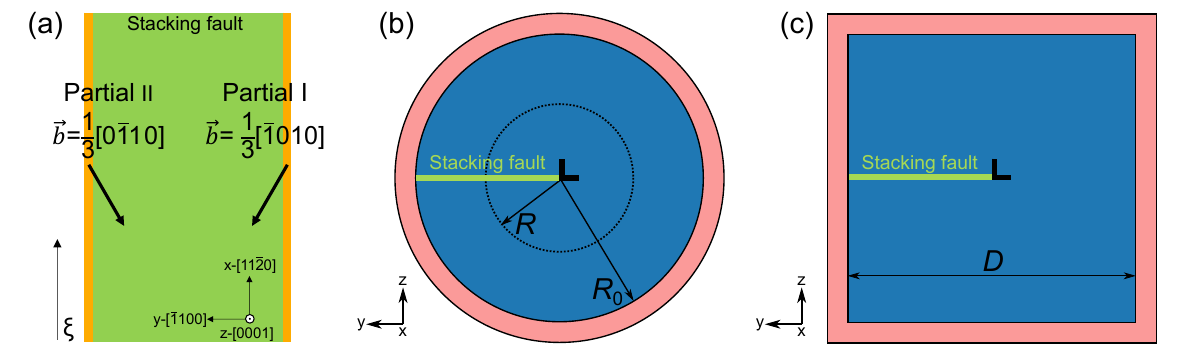}
\caption{(a) Schematic representation of the dissociated  basal screw $\langle a \rangle$ dislocation in hexagonal Laves phases into two 30{\textdegree} synchro-Shockley dislocations: partial I and II. (b) Cylindrical setup for the structural optimization of dislocation cores and the calculation of dislocation core energies. The radius of the cylindrical setup ($R_\text{0}$) is around 1000 \AA. (c) Slab setup for the nudged elastic band (NEB) calculation of minimum energy path (MEP) of dislocation motion. 
The dimensions of the slab ($D$) in non-periodic directions are around 300 \AA. 
Periodic boundary conditions are applied along the dislocation line direction ($\xi$) in both setups.
Semi-fixed outer layers where atoms are frozen in non-periodic directions are marked in light red, and the thickness is more than 2 times the interatomic potential cutoff ($>$ 14 \AA).
Please note that the shown setup corresponds to partial I, for partial II the stacking fault is on the right of the dislocation.
}
\label{fig0}
\end{figure*}

To obtain the structures for the further study of synchro-Shockley partial dislocations, perfect screw  dislocations with Burgers vectors \textbf{b}\textsubscript{C14} = $a_{0}^\text{C14}$/3 $[ \bar{1} \bar{1} 2 0 ]$ and \textbf{b}\textsubscript{C15} = $a_{0}^\text{C15}$/2 $[ \bar{1} 1 0 ]$ were introduced following the method detailed in \cite{vaid2022pinning}. After relaxation using the conjugate gradient (CG) with box relaxation and the FIRE \cite{bitzek2006structural,guenole2020assessment} algorithm (force tolerance: $10^{-8}$ eV/$\text{\AA}$), the inserted full screw dislocation dissociated into two widely separated 30\textdegree{} synchro-Shockley dislocations with Burgers vectors \textbf{b}\textsubscript{C14} =  $a_{0}^\text{C14}$/3 $[ \bar{1} 0 1 0 ]$ and $a_{0}^\text{C14}$/3 $[ 0 \bar{1} 1 0 ]$ (for C15 CaAl\textsubscript{2}: \textbf{b}\textsubscript{C15} = $a_{0}^\text{C15}$/6 $[ \bar{1} 2 \bar{1} ]$ and $a_{0}^\text{C15}$/6 $[ \bar{2} 1 1 ]$) bounded by stacking faults, see the sketch in FIG. \ref{fig0}(a). 
In the following, the partial dislocation on the right with the edge Burgers vector component along $[ \bar{1} 1 0 0]$ (or $[1 1 \bar{2}]$ for C15 CaAl\textsubscript{2}) direction is termed partial I and the partial dislocation on the left with the edge Burgers vector component along $[ 1 \bar{1} 0 0]$ (or $[\bar{1} \bar{1} 2]$ for C15 CaAl\textsubscript{2}) direction is called partial II.
To investigate partial I and II dislocations separately, the atomic displacement field corresponding to the partial Burgers vector was imposed to the upper half of the crystal ($> l_z/2$) from the end to the center of the simulation box.
After relaxation with the above-mentioned algorithms, partial I and II located at the center of the simulation box with a stacking fault bounded to the box end were obtained. 

To calculate the core energies of the 30\textdegree{} synchro-Shockley dislocations, cylindrical samples were cut out from the initial simulation box ($l_y$, $l_z\approx$ 2000 \text{\AA}) with each  of the 30\textdegree{} partial dislocations located in the center of the simulation setup, and the stacking fault  bounded at the surface of the cylinder as shown in FIG. \ref{fig0}(b). 
Atoms in the outermost layers of the setups with a thickness of 14 \text{\AA} (2 times the interatomic potential cutoff) were fixed in $y$ and $z$ directions. The radius of the cylindrical setup is around 1000 \text{\AA} to reduce the effect of the boundary conditions. Periodic boundary conditions (PBC) were applied in $x$ direction (the direction of the dislocation line). For C14 CaMg\textsubscript{2}, $l_x\approx 18.4$ \text{\AA} and contains 3 unit cells, for C15 CaAl\textsubscript{2}, $l_x\approx 23.6$ \text{\AA} and contains 2 unit cells. In both cases, $l_x$ is more than 2 times larger than the interatomic potential cutoff.

After relaxation with the aforementioned boundary conditions, the core energies were calculated by measuring the total dislocation energy as a function of radius $R$ and then extrapolating the far-field elastic energy back to the chosen cutoff radius ($r_{c}=b$):
\begin{equation}
\label{e1}
E_\text{tot}(R)-N(R)E_\text{0}=K\text{ln}(R/r_{c})+E_\text{SF}(R)+E_\text{core}|r_{c},
\end{equation}
where $E_\text{tot}$($R$) is the energy of the $N$ atoms contained within a cylinder  of radius $R$, $E_\text{0}$ is the atomic cohesive energy of the given composition, $K$ is an elasticity coefficient containing anisotropic elastic constants \cite{anderson2017theory} and the Burgers vector $b$, $E_\text{SF}(R)$ is the energy contribution of the stacking fault as a function of radius $R$, $E_\text{core}|r_{c}$ is the core energy defined at the chosen cutoff radius $r_{c}=b$.
Note that as the stacking fault extends all the way to the edge of the simulation setup, $E_\text{SF}(R)$ is continuously increasing with $R$.
Finally, the excess in strain energy $E_\text{ESE}$ due to the elastic distortion of the lattice induced by the dislocation, normalized with the dislocation line length \textit{L} can be expressed from EQU.~\ref{e2}:
\begin{equation}
\label{e2} 
E_\text{ESE}(R) = \frac{E_\text{tot}(R)-N(R)E_\text{0}}{L}.
\end{equation}
$E_\text{ESE}$ was calculated at 100 different $R$ values from 1$b$ to 250$b$.

Climbing image nudged elastic band (NEB) \cite{henkelman2000climbing,henkelman2000improved} calculations were performed on initial (before dislocation motion) and final (after dislocation motion) atomistic configurations to find saddle points and minimum energy paths (MEPs) of dislocation motion. The initial configurations were built with the 30\textdegree{} partial dislocations located in the center of the slab setups, as illustrated in FIG. \ref{fig0}(c). 
By altering the width of the displacement field for the inserted partial Burgers vector applied on the upper half of the crystal (see \cite{vaid2022pinning} for details), the final configuration contains the same partial dislocation sitting at the next Peierls valley adjacent to the initial one, corresponding to dislocation motion and expansion of stacking fault by one Burgers vector. 
Atoms in the outermost layers of the setups with a thickness of 14 \text{\AA} were fixed in $y$ and $z$ directions. The dimensions of the slab in $y$ and $z$ directions $l_y$ and $l_z\approx$ 300 \text{\AA}. Slab setups with different dislocation line lengths (in PBC) from 3$\times$ to 30$\times$ unit cells ($l_x$ from 18.4 to 184.2 \text{\AA} for C14 CaMg\textsubscript{2}) were simulated. 

The spring constants for parallel and perpendicular nudging forces are both 1.0 eV/$\text{\AA}^{2}$ \cite{MARAS201613}. Quickmin \cite{sheppard2008optimization} was used as the damped dynamics minimizer to minimize the energies across all replicas with the force tolerance of 0.01 eV/$\text{\AA}$. Different numbers of intermediate replicas from 48 to 144 were simulated and all intermediate replicas were initially equally spaced along the reaction coordinate (RC). The first (RC:0) and last (RC:1) reaction coordinates were determined as the configurations located at the local energy minima along the MEP close to the initial and final configurations, respectively, and were fully relaxed before the NEB calculations.

An atomic structural analysis method for Laves phases, Laves phase crystal analysis (LaCA) \cite{xie2021laves}, was used to analyze the dislocation and stacking fault structures in C14 and C15 Laves phases. LaCA can be seen as an advanced common neighbor analysis (CNA) method \cite{Honeycutt1987}, which combines identification of Frank–Kasper clusters with centrosymmetry to unambiguously classified atomic arrangements in Laves crystals \cite{xie2021laves}. For visualization and analysis of atomistic simulations, OVITO \cite{stukowski2009visualization} was used. 

\section{\label{Results}Results}

\subsection{\label{Results1}Dislocation core energies and structures}

\begin{figure*}[htbp!]
\centering
\includegraphics[width=\textwidth]{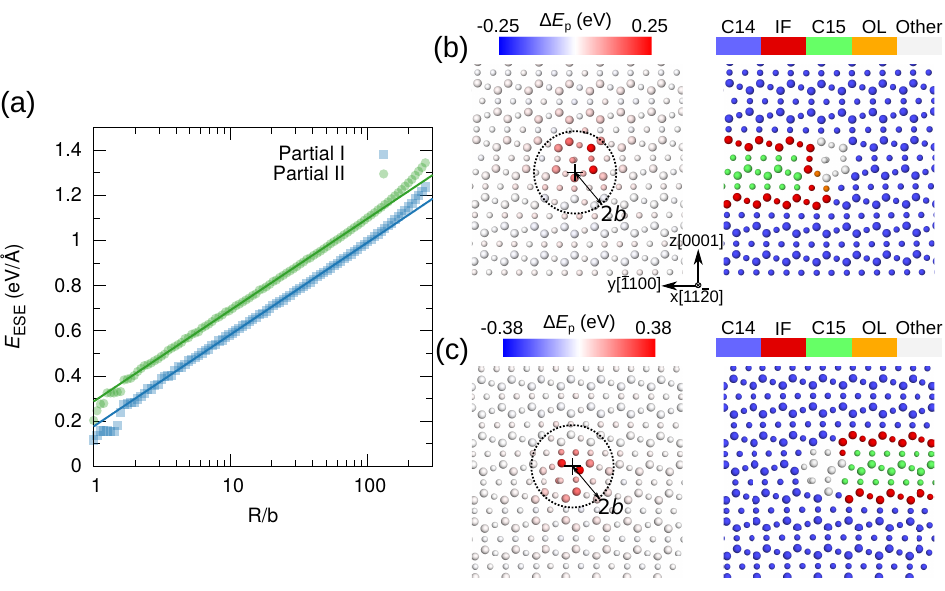}
\caption{(a) Excess in elastic strain  energy $E_{ESE}$ for 30{\textdegree} synchro-Shockley partial I (\textbf{\textit{b}}=$\frac{1}{3}$$[ \bar{1} 0 1 0]$) and II (\textbf{\textit{b}}=$\frac{1}{3}$$[ 0 \bar{1} 1 0 ]$) dislocations in C14 CaMg\textsubscript{2}. The core energy is obtained by extrapolating the far-field elastic energy back to the chosen cutoff radius $r_{c}=b$ ($R/b = 1$, solid lines). Dislocation core structures of 30{\textdegree} synchro-Shockley (b) partial I and (c) II dislocations. Left: colored by deviation of potential energy to bulk ($\Delta E\textsubscript{p}$). Right: colored by Laves phase Crystal Analysis (LaCA). Large and small atoms are Ca and Mg atoms, respectively.}
\label{fig1}
\end{figure*}

Core structures of 30\textdegree{} synchro-Shockley dislocations were analyzed and the corresponding core energies were calculated according to EQU. \eqref{e1}. Two types of 30\textdegree{} synchro-Shockley dislocations with Burgers vectors of $\frac{1}{3}$$[ \bar{1} 0 1 0]$ (partial I) and $\frac{1}{3}$$[ 0 \bar{1} 1 0 ]$ (partial II) were obtained after the energy minimization of the perfect screw dislocations. Both core structures of partial I (FIG. \ref{fig1}(b)) and II (FIG. \ref{fig1}(c)) were observed in Laves crystal structures experimentally \cite{chisholm2005dislocations,zhang2011undulating,cheng2021atomic}. Dislocation core energies of partial I and partial II were calculated by extra\-polating the far-field elastic energy back to the chosen cutoff radius $b$, see FIG. \ref{fig1}(a). The excess in strain energy $E_\text{ESE}$ was plotted against logarithm of the ratio of $R/b$ in FIG. \ref{fig1}(a). The deviation of potential energy within the radius $R<2b$ is significant as shown in FIG. \ref{fig1}(b-c) (and FIG. S1(b-c) in \cite{SuppMat}). In contrast, the atoms belonging to the stacking faults show less energy deviation due to the low stacking fault energies of the Laves phases ($\gamma_\text{SF}^\text{CaMg\textsubscript{2}}$=14 mJ/$\text{m}^\text{2}$ and $\gamma_\text{SF}^\text{CaAl\textsubscript{2}}$=52 mJ/$\text{m}^\text{2}$). The slope of the total energy $E_\text{tot}$ versus ln($R/b$) is close to linear for radii $R$ significantly larger than the core region except when $R$ approaches the semi-fixed boundary, which is in agreement with elasticity theory. A linear model was fitted to the data of $E_\text{ESE}$ vs. $\ln R/b$ from 5$b$ to 100$b$. An elasticity coefficient ($K$/($b^\text{2}/4\pi$)=27.8 GPa) was obtained, which is less than 2\% deviation from the theoretical $K$ value ($K_\text{elast}^\text{30\textdegree}$/($b^\text{2}/4\pi$)=27.3 GPa) calculated considering basal plane isotropy in hexagonal crystals \cite{anderson2017theory} and using the elastic constants of the interatomic potential (see TABLE S I in \cite{SuppMat}):
\begin{equation}
\label{e3} 
K_\text{elast}^\theta = K_\text{elast}^\text{0\textdegree}\text{cos}^2\theta+K_\text{elast}^\text{90\textdegree}\text{sin}^2\theta,
\end{equation}

\begin{equation}
\label{e4} 
K_\text{elast}^\text{0\textdegree} = \frac{ b^2}{4\pi}(C_{44}C_{66})^\frac{1}{2},
\end{equation}

\begin{equation}
\label{e5} 
K_\text{elast}^\text{90\textdegree} = \frac{ b^2}{4\pi}(\bar{C}_{11} + C_{13}) \left[\frac{C_{44}(\bar{C}_{11}-C_{13})}{C_{33}(\bar{C}_{11}+C_{13}+2C_{44})}\right]^\frac{1}{2},
\end{equation}

\begin{equation}
\label{e6} 
\bar{C}_{11}=(C_{11}C_{33})^\frac{1}{2}. 
\end{equation}

For the $\theta$=30\textdegree{} synchro-Shockley dislocation in the simulated C14 CaMg\textsubscript{2}, the elasticity coefficient of the screw component $K_\text{elast}^\text{0\textdegree}$/($b^\text{2}/4\pi$) is 25.6 GPa and the value of the edge component $k_\text{elast}^\text{90\textdegree}$/($b^\text{2}/4\pi$) is 32.6 GPa.

The core energies of partial I and II dislocations at the chosen cutoff radius $r_{c}=b$ are 0.16 and 0.28 eV/$\text{\AA}$, respectively. The core energy of partial II is 75 \% higher than partial I, which indicates that partial I is more energetically favorable than partial II in the simulated C14 CaMg\textsubscript{2} phase. Similar results were also obtained in the simulated C15 CaAl\textsubscript{2} phase as shown in FIG. S1(a) in \cite{SuppMat}.

\subsection{\label{Results2}Mechanisms of dislocation motion}

\begin{figure*}[htbp!]
\centering
\includegraphics[width=\textwidth]{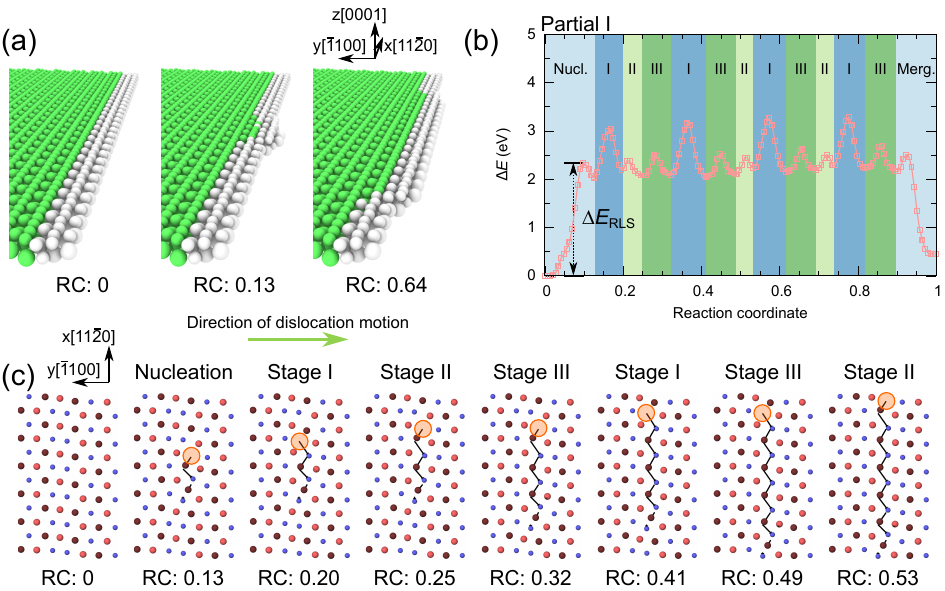}
\caption{Transition mechanism of 30{\textdegree} synchro-Shockley partial I dislocation motion along the $[ 1 \bar{1} 0 0 ]$ direction (i.e., negative $y$ direction, see the coordinate system) via kink propagation in C14 CaMg\textsubscript{2} ($l_x$=6.1 nm, $l_y$=$l_z$=30 nm). (a) Atomic configurations of kink nucleation and propagation. Only atoms belong to stacking fault (C15, colored in green) and dislocation core (others, colored in white) according to LaCA are shown here. (b) Excess energy versus reaction coordinate (RC) is calculated using NEB. Energy profile is separated into different stages based on individual activation events. The activation energy of the rate-limiting step $\Delta E_\text{RLS}$ is marked in the plot. (c) Mechanisms of kink nucleation and propagation. Only atoms in the triple-layer where the dislocation glided are shown here. Large and small atoms are Ca and Mg atoms, respectively. Dark and light red atoms indicate Ca atoms in different layers of the triple-layer. Black arrows indicate displacement vectors (RC: 0 configuration as the reference). Orange circles indicate the locations of vacancies. Green arrow indicates the direction of dislocation motion.}
\label{fig2}
\end{figure*}

The mechanisms of synchro-Shockley dislocation motion were investigated by exploring the transition states between two dislocation structures at adjacent Peierls valleys along the MEP using the NEB method. Overall, we investigated the mechanisms of motion for both partial I and II and also the effect of point defects on the motion of partial I, where point defects were found to form as part of the dislocation motion. 

\subsubsection{Motion of partial I}
Partial I in C14 CaMg\textsubscript{2} exhibits a transition mechanism of dislocation motion via kink-pair nucleation and kink propagation, as shown in FIG. \ref{fig2}(a). The bow-out of a kink-pair occurs around the reaction coordinate (RC) 0.13 with a height of the edge component of the partial Burgers vector, which corresponds to the motion of the dislocation from one Peierls valley to the next. Then the two kinks propagate in opposite directions and finally merge with each other due to the PBC along the dislocation line direction. The motion of partial I is along the $[ 1 \bar{1} 0 0 ]$ direction. The energy profile of the transition processes of kink-pair nucleation and kink propagation is shown in FIG. \ref{fig2}(b). Transition state peaks and intermediates with similar shapes and values in the energy profile appear repeatedly, which indicates similar events are repeatedly activated along the MEP. The transition mechanism of partial I  motion can be divided into several stages: Nucleation (Nucl.), I, II, III, and merge (Merg.). The detailed mechanism of each stage is illustrated in FIG. \ref{fig2}(c) via the atomic displacement relative to the initial configuration (RC 0). To better visualize the individual events between the transition states, only atoms in the triple-layer where the dislocation glides are shown since most atomic movements occurred within the triple-layer. The Ca atoms in different layers within the triple-layer are colored with different shades of red.

\begin{table*}[!htbp]
\centering
\caption[]{\label{tab.1}Activation energies of overall and individual events of the motion of synchro-Shockley dislocations in C14 CaMg\textsubscript{2} ($l_x$=6.1 nm, $l_y$=$l_z$=30 nm). V\textsubscript{X}$\bot$: vacancy at X site at dislocation; X(Y)$\bot$: antisite defect X at Y site at dislocation.}
\centering
\scriptsize
\begin{tabular}{p{0.2\textwidth}p{0.1\textwidth}p{0.1\textwidth}p{0.1\textwidth}p{0.1\textwidth}p{0.1\textwidth}p{0.1\textwidth}}
\hline\hline
\addlinespace[0.1cm]
\multicolumn{1}{l}{} & \multicolumn{6}{c}{Activation energy (eV)}\\
\cmidrule(lr){2-7}
Sample & Overall & Nucleation & Stage I & Stage II & Stage III & Merge \\
\addlinespace[0.1cm]
\hline
\addlinespace[0.1cm]
Partial I & 3.28 & 2.33 & 1.03 & 0.29 & 0.46 & 0.35 \\
Partial I (V\textsubscript{Mg}$\bot$) & 1.53 & - & 1.01 & 0.24 & - & - \\
Partial I (V\textsubscript{Ca}$\bot$) & 1.40 & - & 0.99 & 0.26 & - & - \\
Partial I (Mg(Ca)$\bot$) & 2.59 & 1.75 & 1.03 & 0.21 & - & 0.39 \\
Partial I (Ca(Mg)$\bot$) & 3.15 & 2.29 & 1.08 & - & 0.46 & 0.32 \\
\addlinespace[0.1cm]
\end{tabular}
\begin{tabular}{p{0.2\textwidth}p{0.1\textwidth}p{0.22\textwidth}p{0.15\textwidth}p{0.15\textwidth}}
\hline\hline
\addlinespace[0.1cm]
\multicolumn{1}{l}{} & \multicolumn{4}{c}{Activation energy (eV)}\\
\cmidrule(lr){2-5}
Sample & Overall & Non-sequential shuffling & Shear straining & Rearrangement \\
\addlinespace[0.1cm]
\hline
\addlinespace[0.1cm]
Partial II & 1.79 & 0.31 & 1.10 & 0.11 \\
\hline\hline
\end{tabular}
\end{table*}

In the nucleation stage (from RC 0 to 0.13), a Ca atom (colored dark red) shuffles from top to bottom (the direction along x-$[ 1 1 \bar{2} 0 ]$ is defined as 'up' for ease of reference here) into the lattice and creates a vacancy (marked by an orange circle) and an interstitial in the triple-layer. Meanwhile, the nearby atoms downwards from the Ca atom along the dislocation line shuffle together. The nucleation of the kink-pair can be treated as the formation of two kinks. The formation of the upper and lower kinks are associated with the formation of vacancy and interstitial defects, respectively. The energy barrier of the kink-pair nucleation is 2.33 eV, which is the highest activation energy among all individual events along the MEP of the motion of partial I. Therefore, the kink-pair nucleation process is the rate-limiting step on the reaction coordinate diagram. The energy barriers of overall and individual events along the MEP of dislocation motion in this work are summarized in TABLE \ref{tab.1}. In stage I (from RC 0.13 to 0.20), the Mg atom above the Ca vacancy shuffles to the vacancy and creates another Mg vacancy. This process has an activation energy of around 1.03 eV. Following stage I, stage II (from RC 0.20 to 0.25) corresponds to the shuffling of a Ca atom (colored dark red) to the Mg vacancy and the creation of a Ca vacancy. The activation energy of this event is around 0.29 eV, which is the lowest activation energy of the individual events along the MEP. The combination of stage I and II corresponds to the propagation of the upper kink from bottom to top via repeated formation and occupation of vacancies. Stage III corresponds to the propagation of the lower kink from top to bottom via an interstitial-like mechanism with an energy barrier of around 0.46 eV. The Mg and Ca atoms (colored dark red) below the lower kink shuffle simultaneously and create an interstitial defect at the dislocation core region. In the following events of kink propagation, stage I, III and II repeatedly occur until the kinks merge with an activation energy of 0.35 eV. The final state exhibits higher energy than the initial one because of the expansion of the stacking fault after the motion of the partial dislocation. The largest energy barrier to the motion of partial I  is 3.28 eV. In general, for partial I, the mechanism of dislocation motion is kink-pair formation and propagation in which the kinks propagate via two mechanisms, namely, vacancy-hopping and interstitial shuffling, depending on the character of the kinks. 

Similar mechanisms were identified for partial I dislocations with a longer dislocation length $l_x$=18.4 nm (FIG. S2 in \cite{SuppMat}) and different numbers of intermediate replicas (FIG. S3 in \cite{SuppMat}). In contrast, the two-dimensional (2D) setup with $l_x$=1.8 nm is too small to resolve the complex reactions of kink-pair nucleation and propagation. Instead, the synchronous movement of the Ca (colored dark red) and Mg atoms with one saddle point along the MEP is obtained (see FIG. S4 in \cite{SuppMat}). Although the stacking fault energy of the simulated C15 CaAl\textsubscript{2} is significantly higher than for C14 CaMg\textsubscript{2} (see TABLE S I in \cite{SuppMat}), similar kink-pair nucleation and propagation mechanisms were identified (see FIG. S5 in \cite{SuppMat}).

\begin{figure*}[htbp!]
\centering
\includegraphics[width=\textwidth]{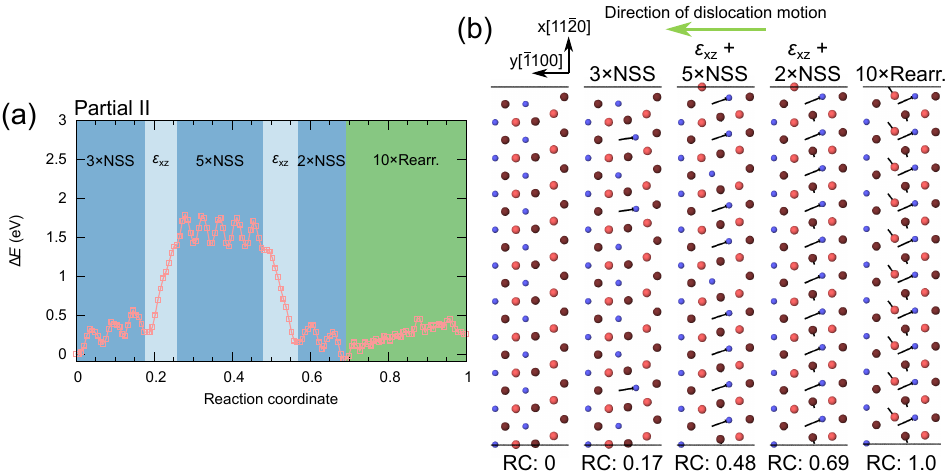}
\caption{Motion of 30{\textdegree} synchro-Shockley partial II dislocation along the $[ \bar{1} 1 0 0 ]$ direction via diffusion-like mechanism in C14 CaMg\textsubscript{2} ($l_x$=6.1 nm, $l_y$=$l_z$=30 nm). (a) Excess energy versus reaction coordinate is calculated using NEB. (b) Diffusion-like mechanism mixed of non-sequential shuffling (NSS), shear straining ($\epsilon_\text{xz}$) and short-range rearrangement (rearr.). Only atoms in the triple-layer where the dislocation glided are shown here. Large and small atoms are Ca and Mg atoms, respectively. Dark and light red atoms indicate Ca atoms in different layers of the triple-layer. Black arrows indicate displacement vectors (RC: 0 configuration as the reference). Green arrow indicates the direction of dislocation motion.}
\label{fig3}
\end{figure*}

\subsubsection{Motion of partial II}
The partial II dislocation shows a different mechanism of dislocation motion compared to partial I. Instead of sequential atomic shuffling during kink propagation of partial I, partial II in C14 CaMg\textsubscript{2} exhibits a non-sequential atomic shuffling during its motion. In addition, the two coupled Shockley partial dislocations of partial II move separately. To investigate the mechanism of motion of leading partial dislocation (extension of stacking fault), the motion of partial II is along the $y$ (\hkl[-1 1 0 0]) direction.

The energy profile and detailed mechanism of partial II dislocation motion are shown in FIG. \ref{fig3}. Similar to the energy profile of partial I, the reaction path of the motion of partial II can be divided into individual events and separated stages, see FIG. \ref{fig3}(a). Between RC 0 and 0.17, three non-adjacent Mg atoms (in the Mg sub-lattice along the dislocation line) shuffle to the adjacent free volume sites separately as shown in FIG. \ref{fig3}(b). Three similar peaks with an average energy barrier of 0.31 eV are obtained for these non-sequential shufflings (NSS). After that, a shear straining ($\epsilon_\text{xz}$) along the $x$ ($[ 1 1 \bar{2} 0 ]$) direction takes place with an activation energy of 1.1 eV and followed by five non-sequential shufflings of Mg atoms (from RC 0.17 to 0.48). The shear straining step has the largest energy difference among all individual events along the MEP of partial II motion. After RC 0.48, an energy drop with the same magnitude as the increase of the energy due to the shear straining occurs because of the release of stored elastic energy. After the non-sequential shuffling of two Mg atoms, the motion of the first of the two coupled Shockley partial dislocations is completed. The motion of the second Shockley partial is carried out by the shuffling of the Ca atoms (colored light red) with an average activation energy of 0.11 eV, which corresponds to an atomic rearrangement (Rearr.) of the dislocation core. The overall energy barrier of partial II dislocation motion is 1.79 eV. 

The 2D setup with $l_x$=1.8 nm exhibits a similar mechanism that also consists of three stages including non-sequential shuffling, shear straining, and atomic rearrangement (see FIG. S6 in \cite{SuppMat}). The activation energy of the shear straining of the 2D setup is 0.33 eV which is proportional to the dislocation length ($l_x$) with the same value per unit length (0.018 eV/\AA) as the three-dimensional (3D) setup ($l_x$=6.1 nm), which indicates the shear straining process is not a localized activation event.

\begin{figure*}[htbp!]
\centering
\includegraphics[width=\textwidth]{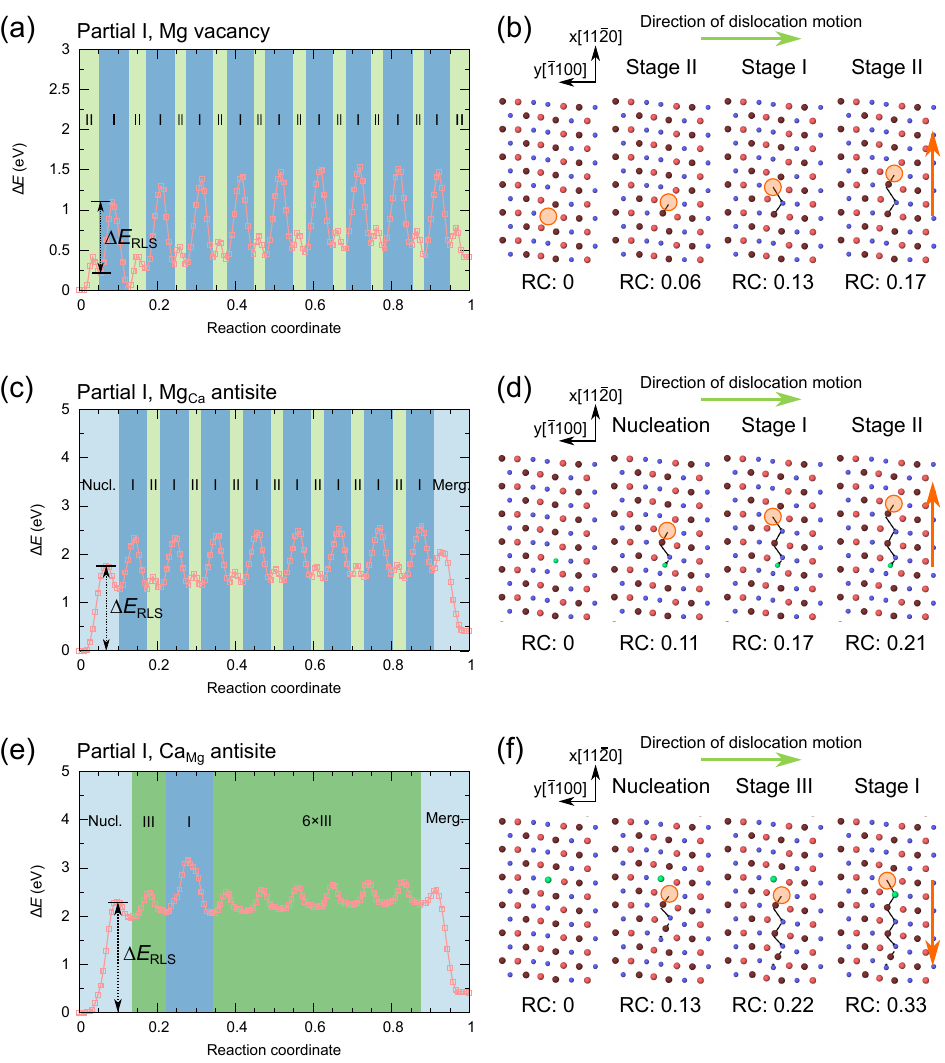}
\caption{Kink propagation of 30{\textdegree} synchro-Shockley partial I dislocation with point defects in C14 CaMg\textsubscript{2} ($l_x$=6.1 nm, $l_y$=$l_z$=30 nm): (a,b) Mg vacancy, (c,d) Mg\textsubscript{Ca} and (e,f) Ca\textsubscript{Mg} antisites. (a,c,e) Excess energy versus reaction coordinate is calculated using NEB. (b,d) Vacancy hopping is the governed mechanism of kink propagation of the dislocations with vacancies and Mg\textsubscript{Ca} antisite. (f) Interstitial-like mechanism governs the kink propagation of the dislocation with Ca\textsubscript{Mg} antisite. Only atoms in the triple-layer where the dislocation glided are shown here. Large and small atoms are Ca and Mg atoms, respectively. Dark and light red atoms indicate Ca atoms in different layers of the triple-layer. Green atoms indicate antisite defects. Black arrows indicate displacement vectors (RC: 0 configuration as the reference). Orange circles indicate the locations of vacancies. Orange and green arrows mark the directions of kink propagation and dislocation motion, respectively.}
\label{fig4}
\end{figure*}

\subsubsection{Effect of point defects on the motion of partial I}
As the motion of partial I is associated with the formation and motion of point defects along the dislocation line, we further investigated the effect of the presence of pre-existing point-defects on the partial dislocation motion. Taking the ordered structure of Laves phases into account, we considered vacancies and anti-site defects on both the Mg and Ca sublattices in the dislocation core region within the triple-layer. The formation energies of these point defects in the simulated Laves phases at the dislocation core region and the bulk counterparts are listed in TABLE S II in \cite{SuppMat}.

The observed mechanisms are presented in FIG. \ref{fig4} (and FIG. S7 in \cite{SuppMat}). The energy profile and atomistic mechanism of motion of partial I with one Mg vacancy are shown in FIG. \ref{fig4}(a,b). To construct atomistic samples with a Mg vacancy, the same Mg atom within the triple-layer along the dislocation line was removed in both initial and final configurations. The energy profile of motion of partial I with the Mg vacancy can be separated into two individual activation events. In contrast to the mechanism of a pristine partial I dislocation (see FIG. \ref{fig2}), no kink nucleation stage is presented as the kink nucleus was introduced by the pre-existing Mg vacancy. Instead, the shuffling of a Ca atom (colored dark red) to the pre-existing Mg vacancy triggers the propagation of the kink (from RC 0 to 0.06). This event is similar to stage II as described in the mechanism of pristine partial I dislocation motion with a comparable activation energy of around 0.24 eV, therefore also named stage II here. From RC 0.06 to 0.13, the Mg atom above the Ca atom along the dislocation line shuffles to the Ca vacancy created after the first stage II. The activation energy of this event is around 1.01 eV, which is the highest among all transition events and again close to stage I as described in the mechanism of pristine partial I dislocation motion and corresponding to a similar atomic shuffling. Thus this event is also named stage I. In the following reaction path, stage II and I repeatedly occur and the kink only propagates from bottom to top, which is dominated by the vacancy-hopping mechanism. Similarly, in the partial I dislocation motion with a Ca vacancy, stage I (with an average activation energy 0.99 eV) and II (with average activation energy of 0.26 eV) also repeatedly occur, see FIG. S7 in \cite{SuppMat}. The vacancy-hopping mechanism dominates the propagation of the kink from bottom to top.
The values of $\Delta E_\text{RLS}$ of the partial I dislocation motion with the Mg and Ca vacancies are 1.01 and 1.09 eV, respectively. These values are much lower than the rate limiting energy barrier of the pristine counterpart.

NEB calculations were also performed on the partial I dislocations with pre-existing anti-site defects at the dislocation core region. For this, a Ca or Mg atom was replaced by a Mg or Ca atom in both initial and final atomistic configurations. FIG. \ref{fig4}(c,d) and FIG. \ref{fig4}(e,f) show the energy profiles and mechanisms of dislocation motion with Mg\textsubscript{Ca} and Ca\textsubscript{Mg} anti-site defects, respectively, where the subscript indicates the original elemental species at this lattice position. In the case of Mg\textsubscript{Ca} anti-site, a kink-pair nucleation event occurs at the beginning of the reaction path (from RC 0 to 0.11), which corresponds to the shufflings of the Mg\textsubscript{Ca} anti-site atom (colored in green) together with a Ca atom (colored dark red) and a Mg atom above the anti-site defect as shown in FIG. \ref{fig4}(d). A Ca vacancy is generated at the upper kink, and a Mg interstitial defect of the anti-site atom is formed at the lower kink. The energy barrier of the kink-pair nucleation in the partial I with the Mg\textsubscript{Ca} anti-site is 1.75 eV, which is the rate limiting step, and the value is lower than the pristine counterpart. In the following reaction path, the upper kink propagates upwards via the vacancy-hopping mechanism, and the lower kink is pinned at the interstitial defect. Stage I and II of the anti-site decorated partial I have similar activation energies (1.03 and 0.21 eV for stages I and II, respectively) and similar atomic shuffling mechanisms to the pristine partial I dislocation. The two stages take place repeatedly in the motion of the partial I dislocation with the Mg\textsubscript{Ca} anti-site. %

Different from the vacancy-hopping dominated mechanism in partial I with pre-existing vacancies and a Mg\textsubscript{Ca} anti-site defect, a partial I with a Ca\textsubscript{Mg} anti-site shows an interstitial-shuffling mechanism of dislocation motion (see FIG. \ref{fig4}(e,f)). The kink-pair nucleation is associated with the shufflings of Ca atoms (colored dark red) and the Mg atom below the Ca\textsubscript{Mg} anti-site (colored green) and the generation of a Ca vacancy. The energy barrier of kink-pair nucleation is 2.29 eV, which is the rate limiting step and close to the value of the pristine counterpart. In the rest of the reaction path, the Ca\textsubscript{Mg} anti-site atom shuffles to the Ca vacancy with an activation energy of 1.08 eV, close to the energy barrier of stage I. Then, the upper kink is pinned at the vacancy. The presence of five Ca atoms surrounding the vacancy leads to a high packing density and therefore a small free volume. As a result, the vacancy-hopping mechanism is no longer energetically favorable. Instead, the lower kink propagates downwards via the interstitial-like mechanism with an average activation energy of 0.46 eV, similar to stage III as described for the pristine counterpart.

\section{\label{Discuss}Discussion}

\subsection{\label{Discuss1}Direction-dependence of plasticity}

Two 30\textdegree{} synchro-Shockley dislocations with edge components of Burgers vectors in opposite directions and different core structures were identified as shown in FIG. \ref{fig0}(a) and FIG. \ref{fig1}(b,c). These partial dislocations, named partial I and II here, possess different core energies, which indicates these two dislocations also have different critical stresses for nucleation. Depending on which partial dislocation is more geometrically (higher Schmid factor) and/or energetically (lower activation energy) favorable, nucleation-controlled plasticity could therefore exhibit directional dependencies. %

Additionally, the two synchro-Shockley dislocations are expected to exhibit different critical resolved shear stresses for motion due to different mechanisms of motion and corresponding activation energies. 
Specifically, partial I and II exhibit different self-pinning characters during the dislocation motion.

The motion of a synchro-Shockley dislocation, also known as a zonal dislocation, can be regarded as the cooperative motion of two coupled Shockley partial dislocations on adjacent planes of the triple-layer.
The motion of partial I dislocations is dominated by the kink-pair propagation and requires consecutive motion of the two coupled Shockley partial dislocations. I.e., in the vacancy-hopping dominant dislocation motion, as given in TABLE \ref{tab.1}, the energy barrier at 0 K for the motion of one of the coupled Shockley dislocations (corresponding to the stage II with an energy barrier ${\Delta E_\text{stage II}}\approx$0.25 eV) is much lower than the other one (corresponding to the stage I with an energy barrier ${\Delta E_\text{stage I}}\approx$1.03 eV). Thus, the thermally activated kink propagation requires the activation of two consecutive thermally activated mechanisms. If one of these is unsuccessful, especially the one with higher activation energy, then the total synchro-Shockley dislocation becomes temporarily immobile, which is referred to as the self-pinning nature of synchro-Shockley dislocations in Laves phases \cite{kazantzis2007mechanical,kazantzis2008self}.

For partial II, the shear straining process has a higher energy barrier than the two thermally activated events (non-sequential shuffling and short-range rearrangement). Thermal activation could only control the motion of partial II if the applied shear strain/stress reaches a certain level to overcome the corresponding energy barrier. In addition, the motion of the two coupled Shockley partial dislocations of partial II occurs separately, namely, the motion of one of the Shockley partial dislocations (corresponding to the series of non-sequential shuffling processes) takes place first and is then followed by the other Shockley partial dislocation (which moves by the series of short-range rearrangement processes). Unlike the self-pinning character of partial I, the thermally activated motion of partial II does not require consecutive thermal activation of events with different energy barriers.

These observations could be investigated experimentally using micro/nanopillar compression. The nucleation-controlled plasticity may be expected to manifest in micro/nanopillar compression tests on defect-free single crystal Laves phases oriented to maximize resolved shear stresses on either partial I or II dislocations. In contrast, variations in the lattice friction and thermal activation of the partial dislocation motion may be accessible by compression of similarly oriented single crystalline pillars but with a pre-existing dislocation density at different rates and temperatures. However, the controlled preparation of such samples is challenging and whether the expected effects will be possible to resolve given the experimental uncertainties and the need to suppress fracture remains to be explored.

\subsection{\label{Discuss2}Point defect assisted dislocation motion}

Point defects, including vacancies and anti-site atoms, very commonly exist in Laves phases and have significant effects on mechanical properties \cite{zhu1999point,stein2021laves}. For the Mg-Al-Ca alloying system, first-principles calculations on stoichiometric C14 CaMg\textsubscript{2} \cite{shao2015native} and C15 CaAl\textsubscript{2} \cite{tian2017first} suggest the predominance of constitutional anti-site and vacancy defects, respectively. 

The dislocation core region is more energetically favorable for the formation of point defects than the perfect and unstrained Laves crystal lattice, as shown in TABLE S II in \cite{SuppMat}. Therefore, the constitutional point defects are likely to locate at pre-existing dislocations and thus affect the mechanisms of dislocation motion and corresponding activation energy. 
In this study, the key mechanism of vacancy-assisted kink propagation, namely vacancy-hopping, is demonstrated by NEB calculations. 
Furthermore, the presence of a vacancy at the dislocation core region dramatically reduces the energy barriers of kink-pair nucleation. In the simulated C14 CaMg\textsubscript{2}, the presence of V\textsubscript{Mg} and V\textsubscript{Ca} lower the activation energies of the rate limiting step $\Delta E_\text{RLS}$ of dislocation motion by 57\% and 58\%, respectively. 

Anti-site defects have also been proposed to affect the hardness of Laves phases with off-stoichiometric compositions \cite{zhu1999point,voss2008composition,takata2016nanoindentation,luo2020composition} and progressive softening behavior with deviations from stoichiometric Laves phases has indeed been observed in previous experiments \cite{voss2008composition,takata2016nanoindentation,luo2020composition}. In this study, a possible origin of this behavior is unveiled by considering the influence of segregated anti-site defects at the dislocation core on the mechanism and activation energy of kink-pair nucleation and propagation. In the simulated C14 CaMg\textsubscript{2} phase, the effect of anti-site defects on the rate limiting activation barrier of dislocation motion is a reduction ranging from 2\% to 25\% depending on the anti-site type. The effect of Mg\textsubscript{Ca} anti-site defects is much more pronounced than Ca\textsubscript{Mg} anti-site defects. The reason for this is that a small Mg atom on a large Ca atom site generates excess free volume, making it easier for the lattice to adapt to the formation of vacancy and interstitial during kink nucleation, thus facilitating the vacancy-hopping mechanism of kink propagation. This finding agrees well with the experimental results that the softening was observed in off-stoichiometric compounds that were rich in the smaller B-atoms \cite{voss2008composition,takata2016nanoindentation}. 

The thermal fluctuation can not only lower the critical stress of dislocation motion but also speed up the atomic diffusion which results in the formation of thermal vacancies \cite{zhu1999point,tian2017first}. At finite temperatures below the temperatures in which diffusion enables diffusion-based mechanisms of motion, a significant number of vacancies will favour vacancy-hopping mechanisms and thus have a prominent effect on the mechanisms of dislocation motion.

How the different activation barriers for the migration of the two kinks affect the overall mobility and resulting shapes of the partial I dislocation for different stresses and temperatures will depend on the multiple factors, including the pinning distance of the dislocations, the types of obstacles for dislocation motion, the mean free path for dislocation motion, the rates of nucleation and annihilation of kinks and the concentration and mobility of point defects. All these factors are outside the scope of this study, but could be addressed by, e.g., kinetic Monte Carlo (kMC) simulations, like in \cite{cai2001kinetic,stukowski2015thermally,shinzato2019atomistically}.
The identified activation events and correlated activation energy barriers determined in this work could serve as input parameters for developing a kMC model of dislocation motion in Laves phases. The kMC model could allow investigations of dislocation dynamics comprising various kinds of events with atomistically informed activation rates.

\subsection{\label{Discuss3}Generalizability of the mechanisms}
The outcomes of this study, namely the mechanisms of direction-dependent and point defect assisted motion of synchro-Shockley dislocations in C14 CaMg\textsubscript{2} and C15 CaAl\textsubscript{2} Laves phases, could be generalized to other Laves phases and crystal structures containing Laves crystal building blocks, such as $\mu$ phases. 

In this work, two distinct core structures and core energies of partial I and II dislocations were identified in C14 CaMg\textsubscript{2} and C15 CaAl\textsubscript{2} Laves phases (see FIG. \ref{fig1}, and FIG. S1 in \cite{SuppMat}). 
Dislocations often have different equivalent and non-equivalent core structures resulting from the symmetry of the crystal lattice \cite{caillard2003thermally}. In Laves crystal structures, the atomic arrangement within the triple-layer is different at two ends of the synchroshear-induced stacking fault along $\langle a \rangle$ or $\langle 1 1 0 \rangle$ direction (see FIG. S8 in \cite{SuppMat}). The surrounding atomic distributions of the smaller B-atoms at the end of the stacking fault (with a Burgers vector of 1/3 $[ \bar{1} 0 1 0 ]$ or 1/6 $[ \bar{1} 2 \bar{1} ]$ corresponding to the extension of the stacking fault relative to the matrix) are non-centrosymmetric (see FIG. S8(e-f) in \cite{SuppMat}). In contrast, the smaller B-atoms at the opposite end of the stacking fault (with the Burgers vector of 1/3 $[ 0 \bar{1} 1 0 ]$ or 1/6 $[ \bar{2} 1 1 ]$) exhibit centrosymmetric atomic environment within the triple-layer. Therefore, partial I and II dislocations bounding the synchroshear-induced stacking fault with 30{\textdegree} Burgers vectors symmetric with respect to the dislocation line direction ($\langle a \rangle$ or $\langle 1 1 0 \rangle$ direction) have different core structures and energies. 
Accordingly, different nucleation criteria and mobilities of partial I and II dislocations are expected to generally exist on the deformation-mechanism maps of Laves crystal building blocks.
Similar observations were reported in diamond-like \cite{koizumi2000core,blumenau2002dislocations} and perovskite structures \cite{gumbsch2001plasticity}, where the dislocations with identical/symmetric Burgers vector have different core structures at low and high temperatures thus exhibiting temperature-dependent mobilities. 

The kink-pair mechanism is a dislocation glide mode across Peierls valleys as a consequence of high lattice friction \cite{caillard2003thermally}. In previous atomistic studies, the nucleation and propagation of kink-pairs were found to control the glide of screw dislocations in body-centred cubic iron \cite{wen2000atomistic,narayanan2014crystal}, silicon \cite{pizzagalli2008theoretical,pizzagalli2016atomistic} and perovskite crystals \cite{goryaeva2016low,kraych2016peierls}, where the activation volume of the kink-pair nucleation is only a few $b^3$. Laves phases are notoriously brittle at low temperature due to the difficulty in moving dislocations under external shear stress \cite{stein2021laves,livingston1992laves}, therefore the kink-pair mechanism is believed to govern the motion of synchro-Shockley dislocations which serve as the carrier of plasticity at high temperature. The nature of kink-pair controlled dislocation motion in the simulated Laves phases correlates well with the experimental estimations on the activation volume of C14 CaMg\textsubscript{2} ($\Omega = 13 b^3$, based on the experimental data obtained from micropillar compression tests \cite{freund2021plastic, zehnder2019plastic} and the calculation details described in \cite{SuppMat}) and other Laves phases \cite{ohba1989high,saka1993plasticity,kazantzis2007mechanical,kazantzis2008self}. 
Furthermore, the effects of vacancy and anti-site defects on the kink-pair mechanism of dislocation motion are believed not to be limited to specific Laves compounds. It was speculated that the presence of a vacancy or anti-site defect at the site of a kink would facilitate the motion of synchro-Shockley dislocations from the geometry and packing density of Laves crystal structure \cite{kumar2004polytypic,takata2016nanoindentation}.

\section{\label{Conclude}Conclusions}

In this study, we investigated the mechanisms of motion of 30\textdegree{} synchro-Shockley dislocations in C14 CaMg\textsubscript{2} and C15 CaAl\textsubscript{2} Laves phases using atomistic simulations. The MEP of dislocation motion and corresponding activation energies were determined using the NEB method. 

Our aim was to reveal the mechanisms of motion of synchro-Shockley partial dislocations in Laves phases and to begin to understand the physical origins of changing mechanical properties of Laves phases containing point defects as a result of temperature and stoichiometry changes. From this work, we conclude that:

\begin{itemize}
  \item Two types of 30\textdegree{} synchro-Shockley dislocations with different Burgers vectors (referred to as partial I and partial II) were identified in the simulated C14 CaMg\textsubscript{2} and C15 CaAl\textsubscript{2} Laves phases and observed experimentally in other Laves or structurally similar phases. Partial I exhibits a lower core energy than partial II, it is therefore expected to have a lower critical nucleation stress.
  \item Partial I dislocation propagates via kink-pair nucleation and propagation. Kink-pair nucleation on partial I dislocations is accomplished by creating a vacancy and interstitial pair. The kink-pair then propagates via  vacancy-hopping and interstitial-shuffling mechanisms in two directions along the dislocation line separately.
  \item The motion of partial II dislocation in C14 CaMg\textsubscript{2} consists of three stages including non-sequential atomic shuffling, shear straining and atomic rearrangement. The shear straining process exhibits the largest energy difference among these individual stages and is not a localized activation event. In contrast, the non-sequential atomic shuffling and atomic rearrangement processes are highly localized events. 
  \item The presence of point defects at the dislocation core significantly lowers the energy barrier for the motion of the partial I dislocation in C14 CaMg\textsubscript{2}. In the cases of vacancies and B\textsubscript{A} anti-site defect, the activation energy of kink-pair nucleation is dramatically reduced and the propagation of one of the kinks is dominated by the vacancy-hopping mechanism.
\end{itemize}

\begin{acknowledgments}
The authors acknowledge financial support by the Deutsche Forschungsgemeinschaft (DFG) through Projects No. A02, No. A05, and No. C02 of the SFB1394 Structural and Chemical Atomic Complexity—From Defect Phase Di- agrams to Material Properties, Project ID 409476157. This project has received funding from the European Research Council under the European Union’s Horizon 2020 research and innovation programme (Grant Agreement No. 852096 FunBlocks). E.B. gratefully acknowledges support from the Deutsche Forschungsgemeinschaft through Project No. C3 of the Collaborative Research Centre SFB/TR 103. Simulations were performed with computing resources granted by RWTH Aachen University under Project No. rwth0591, by the Erlan- gen Regional Computing Center (RRZE) and by the EXPLOR center of the Université de Lorraine and by the GENCI-TGCC (Grant No. 2021-A0100911390 and 2022-A0120911390). We acknowledge Dr. Ali Tehranchi (Max-Planck-Institut für Eisenforschung GmbH) for providing the DFT data. Z.X. would like to thank Dr.-Ing. Wei Luo (RWTH Aachen Uni- versity) for fruitful discussions.
\end{acknowledgments}

\bibliography{main}%

\end{document}

\preprint{APS/123-QED}

\title{Supplementary information: \\Unveiling the mechanisms of motion of synchro-Shockley dislocations in Laves phases}

\author{Zhuocheng Xie}
\affiliation{Institute of Physical Metallurgy and Materials Physics, RWTH Aachen University, 52056 Aachen, Germany}

\author{Dimitri Chauraud}
\affiliation{Max-Planck-Institut für Eisenforschung GmbH, Max-Planck-Str. 1, 40237 Düsseldorf, Germany}

\author{Achraf Atila}
\affiliation{Max-Planck-Institut für Eisenforschung GmbH, Max-Planck-Str. 1, 40237 Düsseldorf, Germany}
\affiliation{Department of Materials Science and Engineering, Institute I: General Materials Properties, Friedrich-Alexander-Universität Erlangen-Nürnberg, 91058 Erlangen, Germany }

\author{Erik Bitzek}
\affiliation{Max-Planck-Institut für Eisenforschung GmbH, Max-Planck-Str. 1, 40237 Düsseldorf, Germany}
\affiliation{Department of Materials Science and Engineering, Institute I: General Materials Properties, Friedrich-Alexander-Universität Erlangen-Nürnberg, 91058 Erlangen, Germany }

\author{Sandra Korte-Kerzel}
\affiliation{Institute of Physical Metallurgy and Materials Physics, RWTH Aachen University, 52056 Aachen, Germany}

\author{Julien Guénolé}
\email[]{julien.guenole@univ-lorraine.fr}
\affiliation{Université de Lorraine, CNRS, Arts et Métiers ParisTech, LEM3, 57070 Metz, France}
\affiliation{Labex Damas, Université de Lorraine, 57070 Metz, France}

\date{\today}

\maketitle

\onecolumngrid

\begin{table*}[!htbp]
\centering
\caption[]{\label{tabS.1}Potential properties of AB\textsubscript{2} Laves phases using the Mg-Al-Ca MEAM potential. $a\textsubscript{0}$ and $c\textsubscript{0}$: lattice parameters; $E_\text{0}$: cohesive energy; $C\textsubscript{ij}$: elastic constants; $B$: isotropic bulk modulus (Hill approximation); $G$: isotropic shear modulus (Hill approximation); $\nu$: Poisson's ratio (Hill approximation); $\gamma\textsubscript{SF}$: basal (\{111\}) stacking fault energy.}
\centering
\scriptsize
\begin{tabular}{l@{\hspace{1cm}}c@{\hspace{1cm}}c@{\hspace{1cm}}c@{\hspace{1cm}}c@{\hspace{1cm}}c@{\hspace{1cm}}c}
\hline\hline
\addlinespace[0.1cm]
\multicolumn{1}{l}{} & \multicolumn{3}{c}{C14 CaMg\textsubscript{2}} & \multicolumn{3}{c}{C15 CaAl\textsubscript{2}}\\
\cmidrule(lr){2-4} \cmidrule(lr){5-7}
\addlinespace[0.1cm]
Properties &  Experiment & DFT & MEAM & Experiment & DFT & MEAM \\
\addlinespace[0.1cm]
\hline
\addlinespace[0.1cm]
$a\textsubscript{0}$ (\text{\AA}) & 6.225 \cite{nowotny1946kristallstrukturen} & 6.230 \cite{arroyave2006intermetallics} & 6.142 & 8.040 \cite{schiltz1974elastic} & 8.052 \cite{yu2009first} & 8.350 \\
$c\textsubscript{0}$ (\text{\AA}) & 10.180 \cite{nowotny1946kristallstrukturen} & 10.070 \cite{arroyave2006intermetallics} & 9.991 & - & - & - \\
$E_\text{0}^\text{A}$ (eV) & - & - & -2.398 & - & - & -4.759 \\
$E_\text{0}^\text{B1}$ (eV) & - & - & -1.487 & - & - & -2.381 \\
$E_\text{0}^\text{B2}$ (eV) & - & - & -1.482 & - & - & - \\
$C\textsubscript{11}$ (GPa) & 61.2 \cite{sumer1962elastic} & 59.5 \cite{ganeshan2009elastic} & 70.9 & 97.0 \cite{schiltz1974elastic} & 105.7 \cite{yu2009first} & 89.2 \\
$C\textsubscript{12}$ (GPa) & 17.6 \cite{sumer1962elastic}& 17.8 \cite{ganeshan2009elastic} & 19.6 & 22.4 \cite{schiltz1974elastic} & 31.1 \cite{yu2009first} & 27.4 \\
$C\textsubscript{13}$ (GPa) & 15.0 \cite{sumer1962elastic} & 12.6 \cite{ganeshan2009elastic} & 19.5 & 22.4 \cite{schiltz1974elastic} & 31.1 \cite{yu2009first} & 27.4 \\
$C\textsubscript{33}$ (GPa) & 65.5 \cite{sumer1962elastic} & 66.0 \cite{ganeshan2009elastic} & 70.3 & 97.0 \cite{schiltz1974elastic} & 105.7 \cite{yu2009first} & 89.2 \\
$C\textsubscript{44}$ (GPa) & 19.3 \cite{sumer1962elastic} & 17.4 \cite{ganeshan2009elastic} & 25.5 & 36.6 \cite{schiltz1974elastic} & 37.3 \cite{yu2009first} & 48.0 \\
$C\textsubscript{66}$ (GPa) & 21.8 \cite{sumer1962elastic} & 20.9 \cite{ganeshan2009elastic} & 25.7 & 36.6 \cite{schiltz1974elastic} & 37.3 \cite{yu2009first} & 48.0 \\
$B$ (GPa) & 31.5 \cite{sumer1962elastic} & 30.1 \cite{ganeshan2009elastic} & 36.6 & 47.3 \cite{schiltz1974elastic} & 56.0 \cite{yu2009first} & 48.0 \\
$G$ (GPa) & 21.3 \cite{sumer1962elastic} & 20.4 \cite{ganeshan2009elastic} & 25.6 & 36.9 \cite{schiltz1974elastic} & 37.3 \cite{yu2009first} & 40.2 \\
$\nu$ & 0.22 \cite{sumer1962elastic} & 0.22 \cite{ganeshan2009elastic} & 0.22 & 0.19 \cite{schiltz1974elastic} & 0.23 \cite{yu2009first} & 0.17 \\
$\gamma$\textsubscript{SF} (mJ/$\text{m}^{2}$) & - & 21 \cite{Tehranchi}  & 14 & - & 98 \cite{Tehranchi} & 52 \\
\hline\hline
\end{tabular}
\end{table*}

\begin{figure*}[htbp!]
\centering
\includegraphics[width=\textwidth]{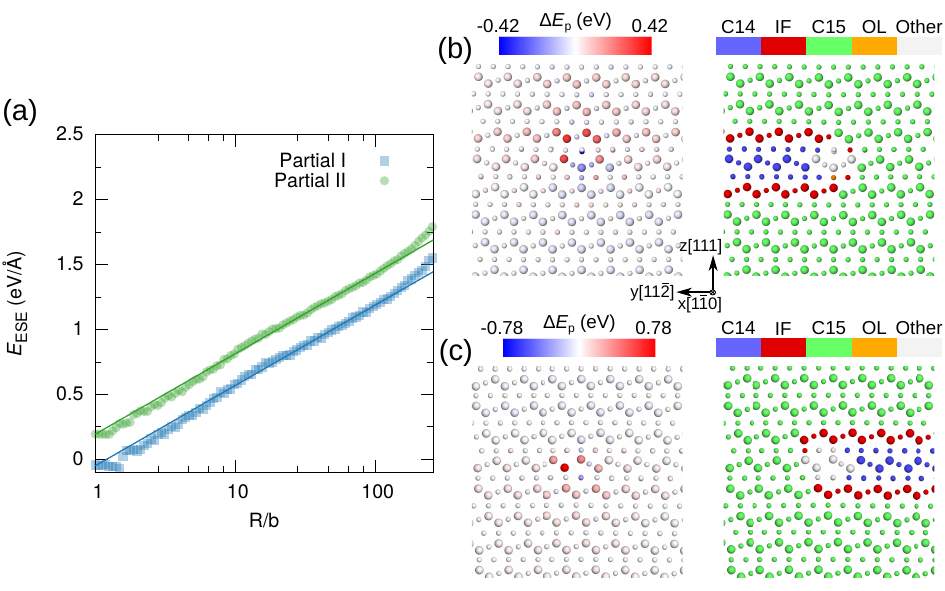}
\caption{(a) Total energy subtracted stacking fault energy for 30{\textdegree} synchro-Shockley partial I and II dislocations in C15 CaAl\textsubscript{2}. The core energy is obtained by extrapolating the far-field elastic energy back to the chosen cutoff radius ($r_{c}=b$). Dislocation core structures of 30{\textdegree} synchro-Shockley (b) partial I and (c) II dislocations. Left: colored by deviation of potential energy to bulk ($\Delta E\textsubscript{p}$). Right: colored by Laves phase Crystal Analysis (LaCA). Large and small atoms are Ca and Al atoms, respectively.}
\end{figure*}

\begin{figure*}[htbp!]
\centering
\includegraphics[width=\textwidth]{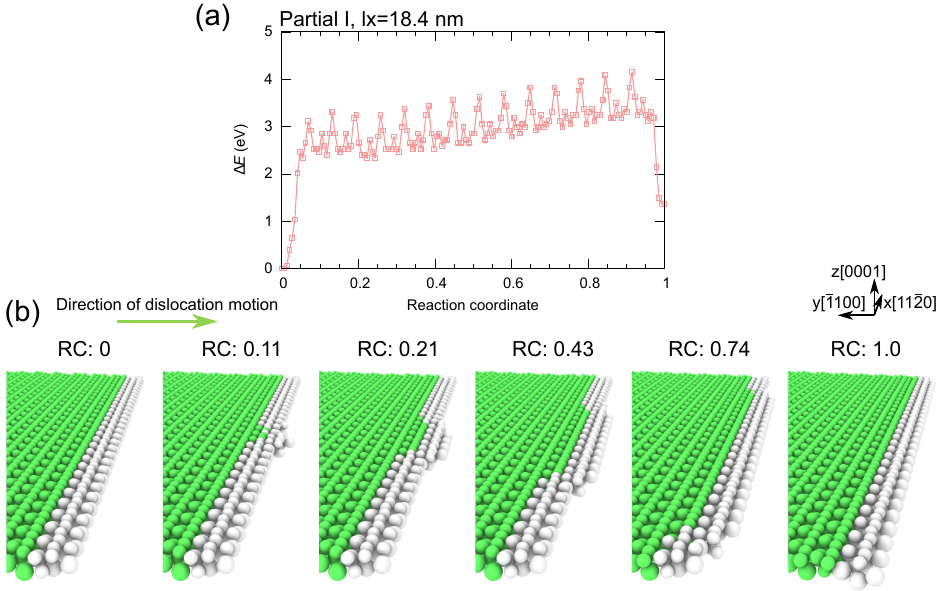}
\caption{Transition mechanism of 30{\textdegree} synchro-Shockley partial I dislocation motion via kink propagation in C14 CaMg\textsubscript{2} ($l_x$=18.4 nm, $l_y$=$l_z$=30 nm). (a) Excess energy versus reaction coordinate (RC) is calculated using NEB. (b) Dislocation motion via kink nucleation and propagation. Only atoms belonging to stacking fault (green) and dislocation core (white) according to LaCA are shown here. Green arrow indicates the direction of dislocation motion.}
\end{figure*}

\begin{figure*}[htbp!]
\centering
\includegraphics[width=\textwidth]{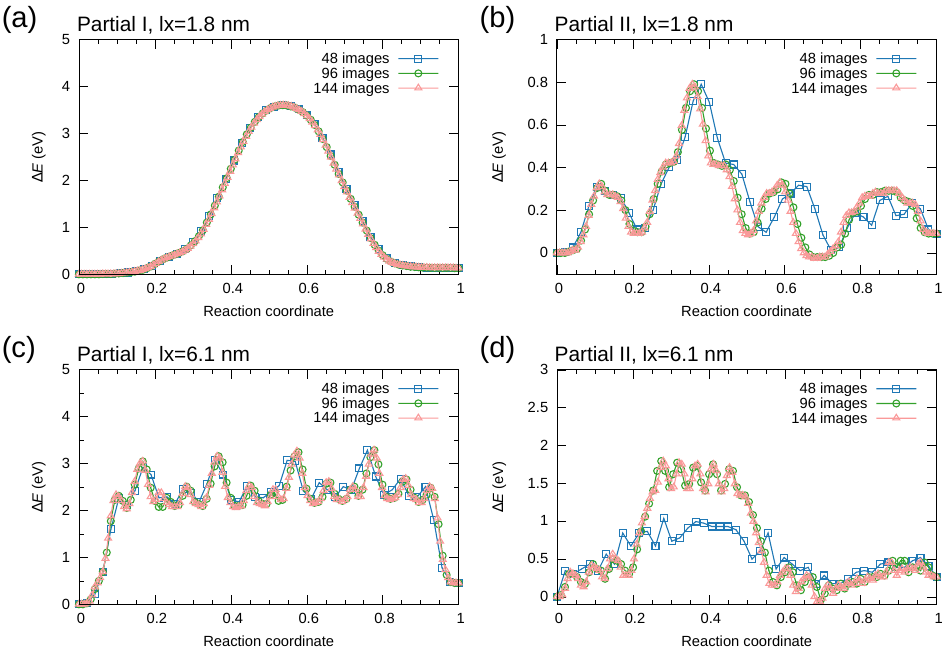}
\caption{Influence of the number of intermediate configurations on the transition mechanism of 30{\textdegree} synchro-Shockley partial I (a,c) and II (b,d) dislocations with different line lengths.}
\end{figure*}

\begin{figure*}[htbp!]
\centering
\includegraphics[width=\textwidth]{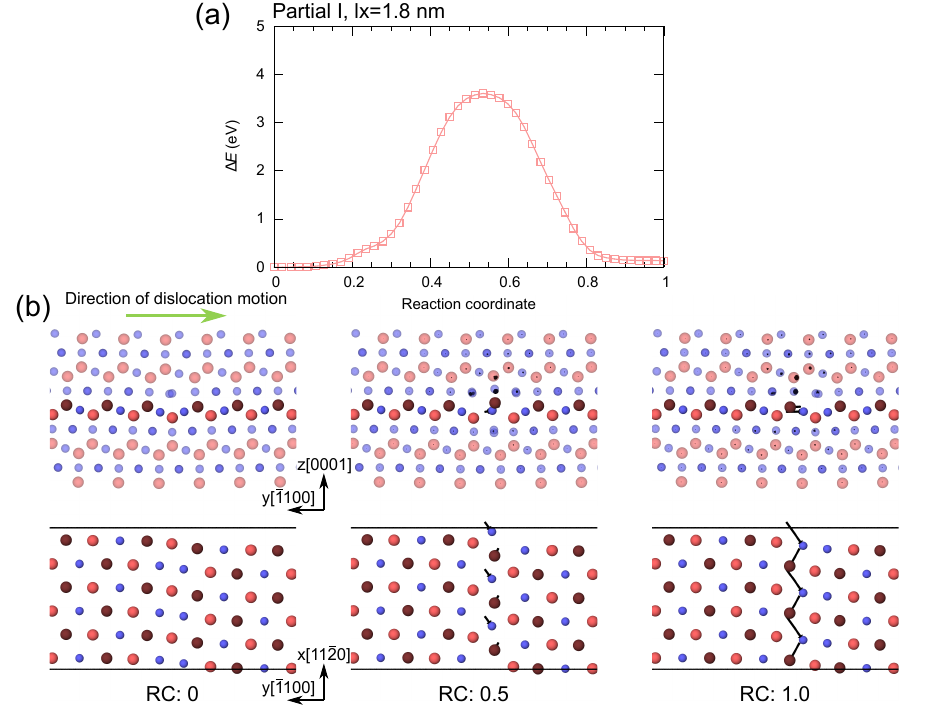}
\caption{Transition mechanism of 30{\textdegree} synchro-Shockley partial I dislocation motion in 2D setup ($l_x$=1.8 nm, $l_y$=$l_z$=30 nm) of C14 CaMg\textsubscript{2}. (a) Excess energy versus reaction coordinate is calculated using NEB. (b) Mechanism of synchro-shuffling of Mg and Ca atoms. Upper: Atoms located outside of the triple-layer where the dislocation glided are half-transparent. Lower: Only atoms in the triple-layer are shown. Large and small atoms are Ca and Mg atoms, respectively. Dark and light red atoms indicate Ca atoms in different layers of the triple-layer. Black arrows indicate displacement vectors (RC: 0 configuration as the reference). Green arrow indicates the direction of dislocation motion.}
\end{figure*}

\begin{figure*}[htbp!]
\centering
\includegraphics[width=\textwidth]{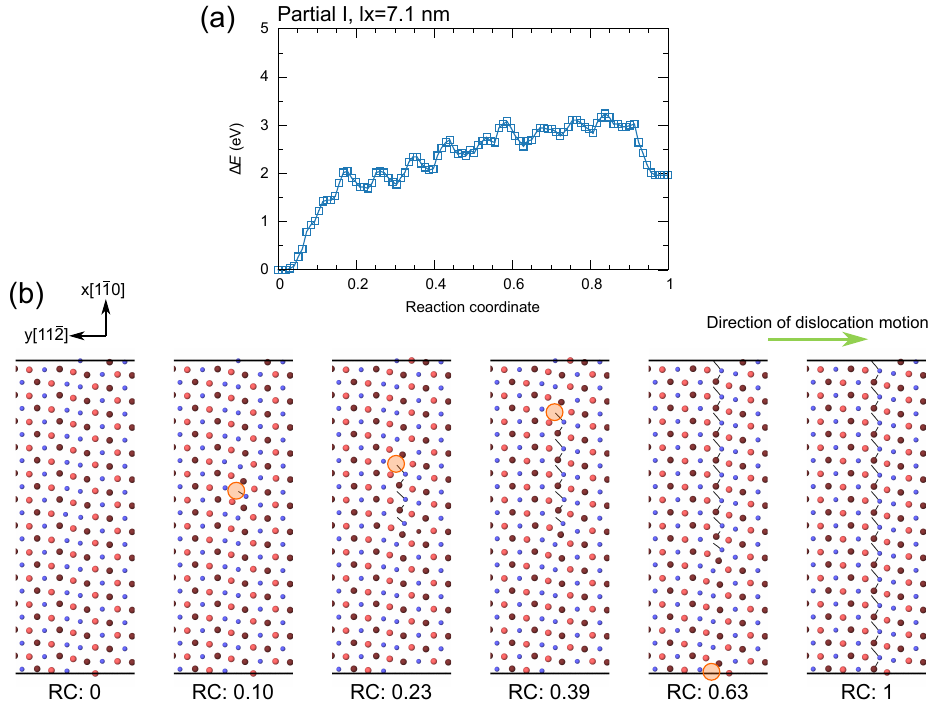}
\caption{Transition mechanism of 30{\textdegree} synchro-Shockley partial I dislocation motion in 3D setup ($l_x$=7.1 nm, $l_y$=$l_z$=30 nm) of C15 CaAl\textsubscript{2}. (a) Excess energy versus reaction coordinate is calculated using NEB. (b) Mechanism of kink nucleation and propagation. Only atoms in the triple-layer where the dislocation glided are shown. Large and small atoms are Ca and Al atoms, respectively. Dark and light red atoms indicate Ca atoms in different layers of the triple-layer. Black arrows indicate displacement vectors (RC: 0 configuration as the reference). Orange circles indicate the locations of vacancies. Green arrow indicates the direction of dislocation motion.}
\end{figure*}

\begin{figure*}[htbp!]
\centering
\includegraphics[width=\textwidth]{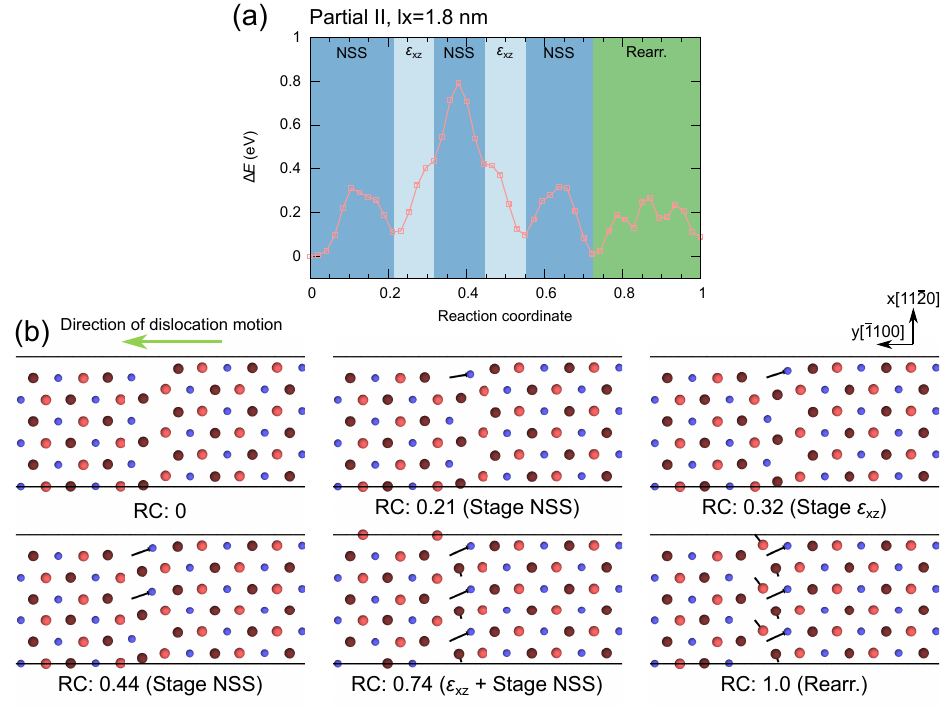}
\caption{Transition mechanism of 30{\textdegree} synchro-Shockley partial II dislocation motion in 2D setup ($l_x$=1.8 nm, $l_y$=$l_z$=30 nm) of C14 CaMg\textsubscript{2}. (a) Excess energy versus reaction coordinate is calculated using NEB. (b) Diffusion-like mechanism of dislocation motion. Only atoms in the triple-layer where the dislocation glided are shown. Large and small atoms are Ca and Mg atoms, respectively. Dark and light red atoms indicate Ca atoms in different layers of the triple-layer. Black arrows indicate displacement vectors (RC: 0 configuration as the reference). Green arrow indicates the direction of dislocation motion.}
\end{figure*}

\begin{figure*}[htbp!]
\centering
\includegraphics[width=\textwidth]{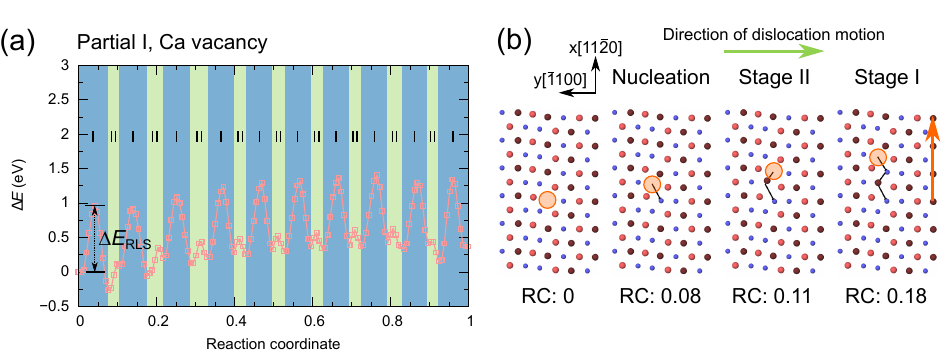}
\caption{Kink propagation of 30{\textdegree} synchro-Shockley partial I dislocation with a Ca vacancy in C14 CaMg\textsubscript{2} ($lx$=6.1 nm, $ly$=$lz$=30 nm). (a) Excess energy versus reaction coordinate is calculated using NEB. (b) Vacancy hopping is the governed mechanism of kink propagation of the dislocations with a Ca vacancy. Only atoms in the triple-layer where the dislocation glided are shown here. Large and small atoms are Ca and Mg atoms, respectively. Dark and light red atoms indicate Ca atoms in different layers of the triple-layer. Black arrows indicate displacement vectors (RC: 0 configuration as the reference). Orange circles indicate the locations of vacancies. Orange and green arrows mark the directions of kink propagation and dislocation motion, respectively.}
\end{figure*}

\begin{figure*}[htbp!]
\centering
\includegraphics[width=\textwidth]{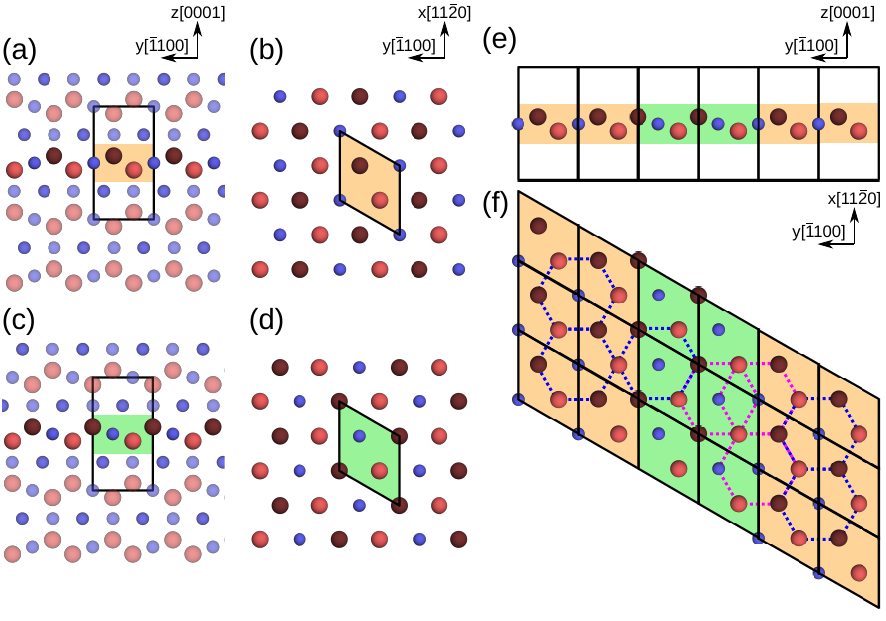}
\caption{Atomic structure of a C14 Laves phase: (a-b) Pristine and (c-d) with a synchroshear-induced stacking fault. (a,c) Side view where atoms located outside of the triple-layer are half-transparent. (b,d) Top view where only atoms in the triple-layer are shown. (e,f) Schematic illustration of the atomic arrangement and lattice pattern across a synchroshear-induced stacking fault (green regions) and matrix (orange regions) in the triple-layer. Dark and light red atoms indicate the larger A-atoms in different layers of the triple-layer. Black boxes show the unit cells. Blue and magenta dashed hexagonal tiles indicate the centrosymmetric and non-centrosymmetric surrounding atomic distributions of the smaller B-atoms, respectively.}
\end{figure*}

\begin{table*}[!htbp]
\centering
\caption[]{\label{tabS.2}Formation energies of point defects in AB\textsubscript{2} Laves phases using the Mg-Al-Ca MEAM potential. V(X): vacancy at X site; X(Y): antisite defect X at Y site; V(X)$\bot$: vacancy at X site at partial I dislocation; X(Y)$\bot$: antisite defect X at Y site at partial I dislocation. Here, $\Delta E\textsubscript{X}$ is the defect formation energy under X-species rich condition, i.e., the chemical potential of the atomic reservoir is zero. Vacancy formation energy: $\Delta E\textsubscript{V(X)}$ = $E\textsubscript{V(X)}$ - $E\textsubscript{tot}$ + $E\textsubscript{X,bulk}$,  $E\textsubscript{tot}$ and $E\textsubscript{V(X)}$ are the total energies of perfect crystal and the crystal contains a vacancy at X site, $E\textsubscript{X,bulk}$ is the energy per atom of X-species elemental solid. Formation energy of antisite: $\Delta E\textsubscript{X(Y)}$ = $E\textsubscript{X(Y)}$ - $E\textsubscript{tot}$ - $E\textsubscript{X,bulk}$ + $E\textsubscript{Y,bulk}$, $E\textsubscript{X(Y)}$ is the total energy of the crystal contains a X(Y) antisite defect.}
\centering
\scriptsize
\begin{tabular}{l@{\hspace{1cm}}c@{\hspace{1cm}}c}
\hline\hline
\addlinespace[0.1cm]
\multicolumn{1}{l}{} & \multicolumn{2}{c}{Formation energy (eV)}\\
\cmidrule(lr){2-3}
Defects &  C14 CaMg\textsubscript{2} & C15 CaAl\textsubscript{2} \\
\addlinespace[0.1cm]
\hline
\addlinespace[0.1cm]
V(B1) & 0.84 & 0.74 \\
V(B2) & 0.86 & - \\
V(A) & 1.47 & 1.48 \\
B(A) & 0.51 & 1.23 \\
A(B1) & 0.42 & 1.00 \\
A(B2) & 0.43 & - \\
V(B)$\bot$ & 0.60 & 0.56 \\
V(A)$\bot$ & 0.98 & 1.24 \\
B(A)$\bot$ & 0.17 & 0.59 \\
A(B)$\bot$ & 0.18 & 0.93 \\
\hline\hline
\end{tabular}
\end{table*}

\clearpage
\subsection{Experimental estimation of activation volume of C14 CaMg$_2$}
The available data on specifically basal slip in the C14 CaMg$_2$ Laves phase studied here allows at least an approximate comparison of the activation volumes \cite{freund2021plastic}. Using the scarce, slip system-specific data with a linear fit through the data points of the critical resolved shear stress from single crystal micropillar compression at room temperature, 150 and 250~\textdegree C \cite{freund2021plastic} (giving a change in stress of 0.09~GPa) and assuming motion of $1/3[10\bar10]$ with $|b|=0.365$ nm, a dislocation density $\rho_m$ of 10$^{12}$ m$^{-2}$, an attempt frequency $\nu_{A}$ of 10$^{11}$~s$^{-1}$, and an average shear strain rate $\dot \gamma$ of 0.001~s$^{-1}$ in 
\begin{equation}
  \Omega=\frac{{{T_1} - {T_2}}}{{{\tau _1} - {\tau _2}}}\frac{k}{1}{\text{ }}\ln \left[ {\frac{{\dot \gamma }}{{{\rho _{\text{m}}}{b^2}{\nu _{\text{A}}}}}} \right]
  , 
\end{equation}
\noindent we find an activation volume of $\Omega = 13 b^3$. This volume is higher than expected for a purely lattice resistance controlled mechanism, however, given the large uncertainty from the underlying experimental data and a possible underestimation of the critical resolved shear stress versus temperature slope due to a lower strain rate used at the lowest temperature \cite{freund2021plastic, zehnder2019plastic}, the calculations in this study and experiments appear consistent at least.

\clearpage
\bibliography{main}%